\documentclass[letters,fleqn,usenatbib]{mnras}

\usepackage[T1]{fontenc}
\usepackage{ae,aecompl}

\usepackage{graphicx}	
\usepackage{amsmath}	
\usepackage{amssymb}	

\usepackage{txfonts} 
\usepackage[utf8]{inputenc} 

\title[A lower limit to the accretion disc radius]{
{A lower limit to the accretion disc radius in the low-luminosity AGN NGC 1052 derived from high-angular resolution data}
}
\author[L. Reb et al.]{%
Lennart Reb,$^{1,2{,3}}$\thanks{E-mail: lennartreb@usm.lmu.de}
Juan A. Fernández-Ontiveros,$^{2,3}$
M. Almudena Prieto,$^{2,3}$
\newauthor 
and Klaus Dolag$^{1,4}$
\\
$^{1}$Universitäts-Sternwarte München, Scheinerstraße 1, D-81679 München, Germany\\
$^{2}$Instituto de Astrofísica de Canarias, C/Vía Láctea, E-38205 La Laguna, Tenerife, Spain\\
$^{3}$Universidad de La Laguna, Dept. Astrofísica, Avd. Astrofísico Fco. Sánchez s/n, E-38206 La Laguna, Tenerife, Spain\\
$^{4}$Max-Planck Institute for Astrophysics, Karl-Schwarzschild-Straße 1, D-85741 Garching, Germany\\
}

\date{Accepted XXX. Received YYY; in original form ZZZ}

\pubyear{2018}

\begin{document}
\label{firstpage}
\pagerange{\pageref{firstpage}--\pageref{lastpage}}
\maketitle

\begin{abstract}
We investigate the central sub-arcsec region of the low-luminosity active galactic nucleus NGC 1052, {using a high-angular resolution dataset that covers 10 orders} of magnitude in frequency.
{This allows} us to infer the continuum emission within {the innermost $\sim {17}\,$pc} around the black hole to be of non-thermal, synchrotron origin and to set a limit to the maximum contribution of a standard accretion disc. 
Assuming the canonical 10 per cent mass-light conversion efficiency for the standard accretion disc, its inferred accretion power {would be} too low by one order of magnitude to account for the observed continuum luminosity.
We thus introduce a truncated accretion disc and derive a truncation radius to mass-light conversion efficiency relation, which we use to reconcile the inferred accretion power with the continuum luminosity.
As a result we find that a truncated disc providing the necessary accretion power must be truncated at $r_\text{tr} \gtrsim {26}\, r_\text{g}$, {consistent} with the inner radius derived from the observations of the Fe K$\alpha$ line in the X-ray spectrum of this nucleus. {This is the first time to derive a limit on the truncation radius of the accretion disc from high-angular resolution data only.}
\end{abstract}



\section{Introduction}

\begin{figure}
\includegraphics[width=\columnwidth]{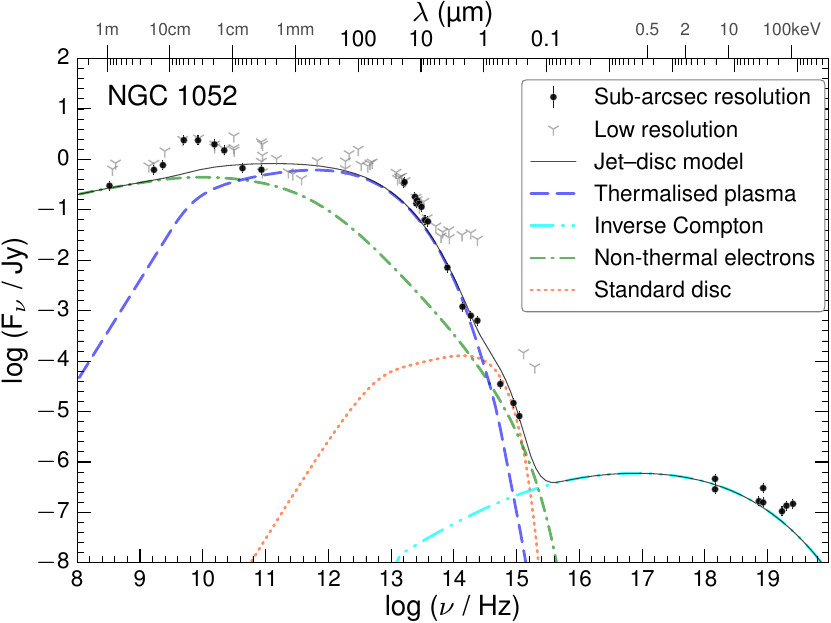}%
\caption[Flux distribution NGC 1052.]{NGC 1052's flux distribution.
Black dots represent the sub-arcsec resolution fluxes ($\lesssim 0\farcs4$), grey spikes are low-angular resolution measurements. The errorbars of the dots (and the size of the spike symbols) span a relative logarithmic error of $\pm 10\,$ per cent.
The compact jet--disc model representation (black thin line) with its individual components: synchrotron emission (blue dashed line) and synchrotron self Comptonisation (cyan double-dot-dashed line) of the thermalised plasma, synchrotron emission of the {non-thermal electrons} (green dash-dotted line), and the maximum contribution of a standard accretion disc (orange dotted line).}%
\label{fig:1}%
\end{figure}

About $\sim 1/3$ of all galaxies in the Local Universe harbour active galactic nuclei (AGNs) with low accretion rates and modest luminosities \citep{2008ARA&A..46..475H}.
These are known as low-luminosity AGNs (LLAGNs) and typically share a number of observational characteristics, which differ from their bright counterparts, Seyfert galaxies and quasars:
they miss the blue bump, i.e. the footprint of the accretion disc, while Fe K$\alpha$ line measurements indicate the absence of optically thick material close to the black hole; the radio loudness, associated to outflows or (compact) jets, tends to increase towards lower luminosities or Eddington ratios \citep{1999ApJ...516..672H,2008ARA&A..46..475H,2005A&A...435..521N}.

The LLAGN in the nearby elliptical galaxy NGC 1052 is an ideal case to study the connection between the accretion flow and the jet emission at low luminosities.
A twin-jet system emanates from the central black hole of $M \sim 1.55\, \times 10^8 \,\text{M}_{\sun}$ \citep{2002ApJ...581L...5W}, with its axis oriented close to the plane of the sky \citep[e.g.][]{2016A&A...593A..47B}.
{It is located} in the southern hemisphere at a distance of $D\sim 18\,$Mpc \citep[$1\arcsec=86\,$pc;][]{2003ApJ...583..712J}, {thus accessible with adaptive optics facilities at the Very Large Telescope (VLT).}
{The use of subarcsec resolution observation}s minimises the contamination {with} host galaxy light in the critical {infrared (IR)} regime, {allowing us to sample the nuclear continuum of the innermost few parsecs.}
X-ray observations {of this source} indicate a nuclear hydrogen column density of $N_\text{H} \sim 10^{22}\, -10^{23}\, \text{cm}^{-2}$ \citep[e.g.][]{2014A&A...569A..26H}.
{A similar column density has been derived from free-free absorption at radio wavelengths, suggesting that ionised gas}
is responsible for the absorption seen in the X-rays \citep[e.g.][]{2004A&A...426..481K}.
Indeed, the silicate emission feature at $9.7\, \micron$ and our parsec-scale dust maps indicate very thin dust around the nucleus with $A_v < 1\,$mag (\citealp{2009MNRAS.397.1966T}; Nadolny et al. in prep.), rendering a free view down to the nucleus.

In this letter we apply a conservative, model independent approach to investigate the power balance in the LLAGN NGC 1052:
we estimate the non-thermal luminosity from the sub-arcsec flux distribution, which we compare with the predictions for the mass flow through the standard accretion disc.
Under the assumption that this mass flow provides all the (accretion) power {required} to supply the non-thermal emission, we investigate {a possible} truncation of the thin accretion disc.

\section{high-angular resolution flux distribution}
\label{sec:sed}
This work is based on sub-arcsec resolution flux measurements of the nuclear region of NGC 1052 in multiple spectral ranges (see \citealp{2012JPhCS.372a2006F,2015ApJ...814..139K}).
In the IR, observations are acquired with the VLT using the Nasmyth adaptive optics system and the near-IR imager and spectrograph (NaCo), {and the} VLT imager and spectrometer for mid-IR (VISIR; \citealt{2014MNRAS.439.1648A}).
In the optical/ultra-violet (UV), nuclear fluxes are measured in archival \textit{Hubble Space Telescope (HST)} images.
All fluxes in the IR and optical/UV range are measured using aperture photometry centred at the unresolved central component and subtracting the local background from a surrounding annulus, and are corrected for galactic reddening ($A_v = 0.073\,$mag; \citealt{2011ApJ...737..103S}).
The apertures are $\sim0\farcs {2}$, radius, except in VISIR ($\sim0\farcs4$).
However, VISIR data {do not have a significant contribution from the host} galaxy \citep{2014MNRAS.439.1648A}. {Both}, high and low-angular resolution measurements are in agreement above $\sim 10\, \micron$, and are thus considered as true {nuclear} emission.

The {continuum flux distribution is} completed with radio measurements, mainly from Very Large Array (VLA) and Very Long Baseline Interferometry (VLBI), and {with} X-ray flux measurements, which have been collected {after} an extensive and careful search in the literature.\footnote{{The high and low-angular flux measurements are available in separate tables on the journal website alongside the paper.}}
Since the contribution of the stellar population is expected to be negligible at high energies,
X-ray fluxes are considered to be of nuclear origin above $2\,$keV and can be compared consistently with sub-arcsec measurements at other frequencies.

The photometric error of the measurements is within a range of $3-5$ per cent, however, we assume a minimum error of 10 per cent on all the measured fluxes to account for possible variability \citep{2005ApJ...625..699M,2005ApJ...627..674A}.

The flux distribution compiled in this way samples the nuclear region over 10 orders of magnitude in frequency at sub-arcsec scales ($\sim 0\farcs {2}$-$0\farcs4$), i.e. it represents the same physical scale around the black hole ($\lesssim 35\,$pc) over nearly the entire accessible electromagnetic spectrum.

\section{Results}
\subsection{The nature of the continuum emission}
The continuum flux distribution shown in Fig.~\ref{fig:1} reveals a constant flux from {radio wavelengths} up to the mid-IR. 
The spectral energy distribution (SED) in Fig.~\ref{fig:2} shows that the mid-IR emission dominates the bolometric luminosity below $100\,$keV.
We isolate a strong point-like source all the way from the mid-IR to the extreme UV, following an inverted {power law} shape over 2 orders of magnitude in frequency ($\alpha \sim 2.58 \pm 0.02$, $F_{1.31\times 10^{13}\,\text{Hz}}\sim0.78\,\text{Jy}$, $F_\nu \propto \nu^{-\alpha}$; \citealt{2015ApJ...814..139K}).
There is no evidence for a noticeable amount of dust extinction in the optical/UV, neither for the contribution of the blue bump, the footprint of the accretion disc.

\citet{2011ApJ...726...87Y} modelled the continuum emission of NGC 1052 with a model {including} a radiatively inefficient accretion flow (RIAF; see \citealt{2014ARA&A..52..529Y} for a review) plus a jet.
The sub-arcsec resolution \mbox{mid-IR} to UV continuum departs largely from the RIAF prediction: the RIAF overestimates the optical/UV flux measurements by more than one order of magnitude {and} underestimates {the mid-IR} flux by roughly one order of magnitude \citep[see fig. 3 in][]{2013EPJWC..6104005F}.

{The observed flat spectrum at radio wavelengths followed by a steep decay at higher frequencies (i.e. flat-inverted) is typically produced by the superposition} of several self-absorbed synchrotron components, as {those produced in} a compact jet \citep{1979ApJ...232...34B}.
\citet{2011ApJ...726...87Y} included a jet in their model to {account for} the radio emission, but the model is incompatible with the IR and optical/UV SED, especially when the new sub-arcsec IR measurements are included.
The reason is that their jet-dominated model assumes $\alpha \sim 0.7$ for the optically thin jet emission, while a much steeper index is required to fit the high-angular resolution data.
A possible explanation is {that the steep IR to UV emission is associated to synchrotron emission by leptons in a thermalised plasma at the base of the jet \citep[e.g.][]{2013MNRAS.434.2696S}. For instance, this configuration can be accomodated by a detailed compact jet model (Fig.~\ref{fig:1}; \citealt{2005ApJ...635.1203M}; Fernández-Ontiveros in prep.).}

In the framework of the Unified Model {for AGNs} \citep{1993ARA&A..31..473A,1995PASP..107..803U} the IR emission is ascribed to thermal emission from dust in a torus, due to the reprocessing of the UV radiation emitted by the accretion disc.
In Fig.~{\ref{fig:2}} the average Seyfert 2 sub-arcsec resolution template of \citet{2010MNRAS.402..724P} is overlaid, {which is based on} a physical region of similar size. 
This flux distribution is dominated by thermal emission from dust and departs significantly from the shape of the IR continuum in NGC 1052.
{Therefore, the IR emission in NGC 1052 does not show a significant contribution from a torus as observed in Seyfert 2 AGNs. 
The detection of $\sim4.5$ per cent polarisation degree in the mid-IR continuum further argues against its thermal origin \citep{1982ApJ...252L..53R}.
}

{The shape of the observed continuum emission does not seem to be a product of thermal or RIAF emission. Instead the flat-inverted spectrum is likely the signature of dominant synchrotron emission from a jet.}
Therefore, we assume that non-thermal mechanisms primarily produce the overall continuum emission and consequently the continuum luminosity.

\subsection{Accretion power and conservative continuum luminosity}
The geometrically thin, optically thick standard accretion disc \citep{1973A&A....24..337S} is used to describe the big blue bump {observed} at high accretion rates.
It is an integral component in the framework of the Unified Model, required to feed the innermost processes.
We model the geometrically thin and optically thick multi colour accretion disc following \citet{1984PASJ...36..741M}. 
The disc emission is independent of the micro-physics and {depends mainly} on the inner edge radius and temperature.
Throughout the letter we fix the disc outer edge to be at $r_\text{out}=2000\,r_\text{g}$, gravitational radius $r_\text{g}=MG \,c^{-2}${, where $G$ is the gravitational constant and $c$ the speed of light,} and the inclination to be $i= 72\degr$ with respect to the line of sight \citep{2004A&A...420..467K, 2016A&A...593A..47B}.
In section~\ref{sec:truncdisc} we will discuss the impact of different inclinations.
We fix the inner edge boundary {of the standard accretion disc} to $r_\text{b}=6\, r_\text{g}$, the innermost stable circular orbit {around a Schwarzschild black hole} \citep{1973A&A....24..337S}.
The maximum contribution of a standard accretion disc matching the optical/UV fluxes imposed by the power law {(Fig.~\ref{fig:2}; \citealt{2015ApJ...814..139K})} is shown in Figs.~\ref{fig:1}~\&~\ref{fig:2} and corresponds to a disc luminosity of $L_\text{SD}\sim 4.0 \times 10^{40}\,\text{erg}\,\text s^{-1}$.

We estimate the maximum accretion {power $P$, i.e. the power gained by accretion through the disc,} by (i) assuming the canonical 10 per cent mass-light conversion efficiency of the standard disc, $L_\text{SD}=\eta_\text{SD}\dot M c^2$, where  {$\dot M$ is the accretion rate and} $\eta_\text{SD}=0.1$, and (ii) that all the accreted mass is available in form of energy, $P=\dot M c^2$.
The latter means that we assume complete mass-energy conversion for the accreted mass and neglect advection into the black hole. {For further simplification we assume a constant $\dot M$ through the accretion process, i.e. we neglect mass-loss due to radiation.}
This yields a mass accretion rate of $\dot M \sim 7.1 \times 10^{-6}\, \text{M}_{\sun} \, \text{yr}^{-1}$ and an accretion power of $P=10\,L_\text{SD}\sim 4.0\times 10^{41}\,\text{erg}\,\text s^{-1}$.
Under the assumption of steady accretion, i.e. no extraction of additional rotational power from the black hole spin, $P$ must account for all subsequent processes, i.e. for the production of the non-thermal continuum luminosity and the jet's kinetic power output.

To estimate the non-thermal continuum luminosity $L_\text{RF}$ we interpolate {a selection of} the measurements using a broken power law, as indicated in Fig.~\ref{fig:2}, integrate {this} flux distribution from radio to UV (${5.26}\times 10^{42}\,\text{erg}\,\text s^{-1} \sim {2.70} \times 10^{-4}\, L_\text{edd}$), and subtract the maximum disc contribution ($2.5 \times 10^{40}\,\text{erg}\,\text s^{-1}$). 
{{We excluded the X-ray regime from the integration, since reprocessing by the inverse Compton effect might result in a double-counting of photons.}
In the integration we included the sub-arcsec resolution measurements and low-angular resolution measurements between $10\,\micron$ and $\sim 3\,$mm, where no sub-arcsec resolution measurements are available.
The low-angular resolution fluxes are consistent with sub-arcsec fluxes at the edges of this range, suggesting that they are dominated by {nuclear} emission.

{This} inferred jet luminosity $L_\text{J}$ might be boosted to a certain extent. 
We calculate the maximum boosting factor for {a} jet inclination of $72\degr$ on the basis of an optically thin spectral index of $\alpha\sim 2.58$ and a conical jet shape, $L_\text{J}=\delta^{2+\alpha}L_\text{RF}$, where the relativistic Doppler factor is defined as $\delta= \gamma^{-1} (1-\beta \cos i)^{-1}$ \citep[e.g.][]{1995PASP..107..803U}.
For the range of intrinsic velocities $0.21 \lesssim \beta \lesssim 0.64$ \citep{2016A&A...593A..47B} and a fixed inclination of $72\degr$ the maximum boosting factor $\delta^{4.58}\sim 1.26$ occurs for $\beta \sim 0.31$.
Thus, we obtain a de-boosted luminosity of $L_\text{RF}\sim {4.2} \times 10^{42}\,\text{erg}\,\text s^{-1}$, which can be considered as a conservative lower limit to the non-thermal rest-frame luminosity. 

\section{Truncation of the accretion disc}
A similar procedure was carried out in a previous work for the LLAGN M87 \citep{2016MNRAS.457.3801P}.
In M87, the maximum luminosity obtained for a standard accretion disc {was} $3.4 \times 10^{41}\,\text{erg}\,\text{s}^{-1}$ {and} the observed continuum luminosity $2.7 \times 10^{42}\,\text{erg}\,\text{s}^{-1}$.
While {in this case} the rest-frame luminosity could be reconciled with the inferred accretion power through a standard {disc, the} jet kinetic power estimated from X-ray cavities was significantly higher ($Q\gtrsim 10^{43}\,\text{erg}\,\text{s}^{-1}$; \citealp[e.g.][]{2006MNRAS.372...21A}).
This comparison cast doubt on the existence of efficient accretion in M87, as it is the case for bright, radiatively efficient AGNs \citep[e.g.][]{2016LNP...905..101M}.

In the case of NGC 1052 the accretion power $P$ of a standard disc is even below the rest-frame luminosity $L_\text{RF}$ -- it is too low by one order of magnitude.
Any hotter standard accretion disc, i.e. with a higher accretion power, violates the spectral limits of the measurements.
Therefore, we investigate in the following the scenario of a truncated accretion disc to explore the possibility of a significantly higher accretion rate with a much lower radiative efficiency.

\subsection{Radiative efficiency of the truncated accretion disc}
We begin with the basic equation defining the spectrum of the classical {\citet{1973A&A....24..337S}} standard accretion disc to derive the mass-light conversion efficiency as a function of the truncation radius $r_\text{tr}$.
The luminosity $dL(r)$ released at distance $r$ from the central mass $M$ per surface unit $dA$ is
\begin{equation}
\frac{dL(r)}{dA}=\frac{3}{8\pi}\dot{M}\frac{GM}{r^3}\left(1-\sqrt{r_\text b/r}\right).
\label{eq:sunyshak}
\end{equation}
The {term $\sqrt{r_\text b/r}$ lowers} the luminosity originating from the innermost regions of the disc, since gravitationally released energy is mechanically transported outwards before being converted into heat \citep{1973A&A....24..337S}.
Integrating the luminosity radiated within an {annulus yields}
\begin{align}
L(r_\text{in},r_\text{out}) &=4\pi \int_{r_\text{in}}^{r_\text{out}} \frac{dL(r)}{dA} \,r \,dr \nonumber \\
&=\frac{3}{2}\dot{M}GM\left[\left(\frac{1}{r_\text{in}}-\frac{1}{r_\text{out}}\right) - \frac{2 \sqrt{r_\text b}}{3}\left(  \frac{1}{r_\text{in}^{3/2}}-\frac{1}{r_\text{out}^{3/2}}  \right)\right].
\end{align}
In order to {derive} the luminosity of a truncated accretion disc we set the inner boundary radius and lower integration limit to the truncation radius $r_\text b=r_\text{in}=r_\text{tr}$.
This is applying the zero-torque condition at the truncation radius, which gives us a lower, hence a more conservative luminosity \citep[][]{1999ApJ...520..298Q}. 
We thus obtain
\begin{equation}
L(r_\text{tr},r_\text{out}) = \frac{\dot{M}GM}{2r_\text{tr}} \left[1-C(r_\text{out}/r_\text{tr})\right],
\label{eq:lowlimcorr}
\end{equation}
where $C(x) =  3x^{-1} -2x^{-3/2}$.
Equalising Eq.~\ref{eq:lowlimcorr} with the direct mass-light conversion $L=\eta_\text{tr} \dot{M}c^2$ yields
\begin{equation}
\eta_\text {tr} = \frac{r_\text{g}}{2 r_\text{tr}} \left[1-C(r_\text{out}/r_\text{tr})\right] \leq \frac{r_\text{g}}{2 r_\text{tr}}.
\label{eq:key}
\end{equation}
{$C(r_\text{out}/r_\text{tr})$ becomes negligible for $r_\text{tr} \ll r_\text{out}$. Therefore, the} mass-light conversion efficiency of the truncated accretion disc $\eta_\text{tr}$, i.e. the fraction of the energy gained through accretion that is radiated by the disc, scales inversely with the truncation radius.

\begin{figure}
\includegraphics[width=\columnwidth]{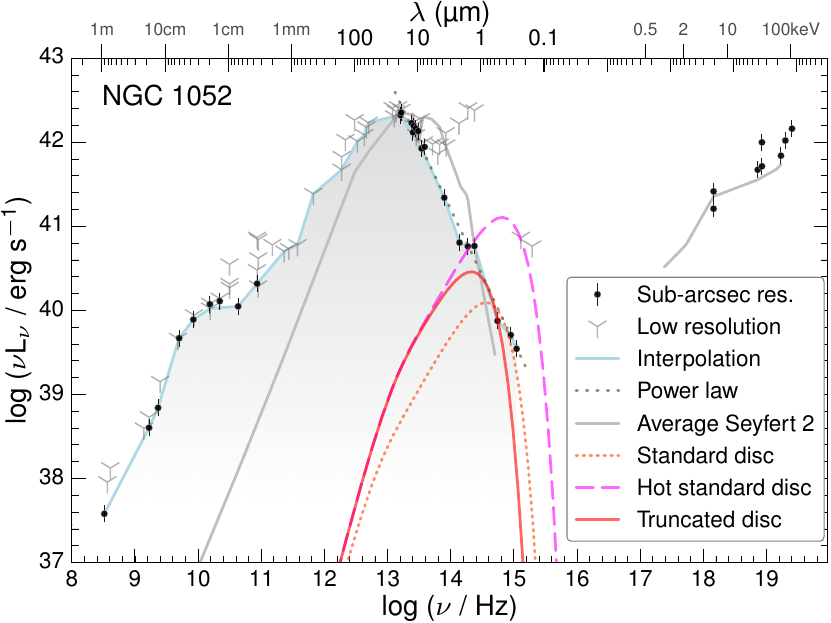}%
\caption[]{NGC 1052's spectral energy distribution (SED). The measurements and the maximum contribution of the standard disc (orange dotted line) {correspond to those shown} in Fig.~\ref{fig:1}. {The broken power law interpolation connecting the selected measurements (light blue line), and the power law with $\alpha=2.58$ (dark-green dotted line) are also shown.
The Seyfert 2 sub-arcsec resolution template of \citet{2010MNRAS.402..724P}, scaled to match the sub-arcsec mid-IR flux, is overlaid (grey line).}
The hot standard disc (magenta dashed line) and the truncated disc (red line) provide each the accretion power {required} to supply the continuum emission.}%
\label{fig:2}%
\end{figure}

\subsection{Constraints to the truncation radius}
\label{sec:truncdisc}
A (hot) standard disc providing the accretion power required to supply the non-thermal continuum luminosity measured, $P= L_\text{RF}$, corresponding to an accretion rate of $\dot M \sim {7.3} \times 10^{-5}\, \text{M}_{\sun} \, \text{yr}^{-1}$, exceeds the limits of optical/UV flux measurements (see Fig.~\ref{fig:2}).
Therefore, the inner hotter part of such a disc must be truncated at a minimum radius to avoid exceeding the continuum flux observed.

To estimate the minimum truncation radius $r_\text{tr}$ we apply an approximate method.
First, following \citet{1999ApJ...520..298Q} the disc luminosity $L$ {depends mainly} on the inner edge temperature $T_\text{in}$ and $r_\text{tr}$, $L\approx 4\pi \sigma r_\text{tr}^2 T_\text{in}^4$.
$L$ is connected to $P$ via the mass-light conversion efficiency $\eta_\text{tr}$.
We adopt $\eta_\text{tr}\propto r_\text{tr}^{-1}$ according to Eq.~\ref{eq:key}, which sets for $P= L_\text{RF}$ a first relation between $r_\text{tr}$ and $T_\text{in}$, $T_\text{in} \propto r_\text{tr}^{-3/4}$. 
Second, the peak of {the} disc spectrum shifts linearly in frequency with $T_\text{in}$ according to Wien's displacement law, and the flux at the peak primarily depends on $r_\text{tr}$ and $T_\text{in}$ as $F_\text{peak}\propto\  r_\text{tr}^2 T_\text{in}^3$ (inserting Wien's displacement law into Planck's law).
We approximate the optical/UV continuum with the power law of \citet{2015ApJ...814..139K}, where we use the (cold) standard disc as a reference point ($T_\text{in}\sim 7400 \,$K).
Equating this flux constraint with $F_\text{peak}$ defines a second scaling relation between both parameters, $T_\text{in} \propto r_\text{tr}^{-2/(3+\alpha)}$.
The intersection of both relations yields a lower limit for $r_\text{tr}$ and an upper limit for $T_\text{in}$.
In other words, this is the solution for a truncated disc of $P= L_\text{RF}$ with a spectrum being `tangent' to the power law continuum observed.
We find that this disc must be truncated at least at $r_\text{tr}\gtrsim {26}\, r_\text{g}$ ($T_\text{in}\lesssim {4400} \,$K, $\eta_\text{tr}\lesssim {2.3}$ per cent),
the corresponding spectrum is shown in Fig.~\ref{fig:2} (red solid line).

{The scaling of $r_\text{tr}$ with $P$ depends on the spectral shape of the measurements, $\log r_\text{tr} \propto (3+\alpha)/(1+3\alpha) \log P\sim 0.64 \,\log P $.}
This means that {in the present case} a disc with a larger truncation radius could supply a substantially higher accretion power without being detected in the SED.
{Including the X-rays ($L_{20-100\,\text{keV}}\sim 1.29 \times 10^{42}\,\text{erg}\,\text s^{-1}$; \citealt{2009A&A...505..417B}) in $L_\text{J}$ would slightly increase the inferred truncation radius to ${r_\text{tr}\gtrsim 30\, r_\text{g}}$, but would not alter our conclusions.} 

We assume in this work an inclination of $72\degr$, which is the highest value of the overlap between the inclination ranges of \citet{2004A&A...426..481K} and \citet{2016A&A...593A..47B}.
For a higher inclination the flux constraints would allow for a hotter standard disc with a higher $P$. For a lower inclination the maximum boosting factor would increase, resulting in a lower $L_\text{RF}$. 
To assess the impact of the geometry in our estimate of $r_\text{tr}$ we explored the range of accessible inclinations $64 \degr \lesssim i \lesssim 87 \degr$ of \citet{2016A&A...593A..47B}.
In particular, we compared the maximum possible accretion power of the standard disc ($i=87\degr$) with the minimum rest-frame luminosity (de-boosted by the maximum possible boosting factor, which occurs for $i=64\degr$). 
Even in this extreme case the accretion power {would be} still lower than the rest-frame luminosity by more than a factor of 2, further arguing against a standard disc transporting the accretion power. 

Possible alternatives to a truncated accretion disc are {the} extraction of rotational energy via the \citet{1977MNRAS.179..433B} mechanism or {the accretion of mass through a geometrically thick and optically thin corona embedding the standard accretion disc} \citep{2001MNRAS.321..759C}. {These mechanisms} could provide additional power to supply the continuum emission.
However, the extraction of rotational energy cannot provide enough power according to current simulations  \citep[e.g.][]{2011MNRAS.418L..79T} {and this process} requires a high $\dot M$ to be efficient  \citep[e.g.][]{2014ARA&A..52..529Y}, which does not seem to be the case for NGC 1052. 
{The optically thin corona} could in principle provide a significant mass flow. In that {case} ${\sim90}$ per cent of the mass flow {would have} to be accreted {through the corona}, which is unlikely.

Previously, an alternative method has been used to claim truncation of the accretion disc in other targets \citep[e.g.][]{1999ApJ...520..298Q,2004ApJ...612..724Y}.
The application of RIAF-based models to the continuum emission yields the mass flow of the RIAF.
Since the spectrum of a standard disc providing this mass flow ($\eta_\text{SD}=0.1$) exceeds the limits of optical/UV flux measurements, the disc must be truncated and is usually assumed to transition into the RIAF at its inner edge.
However, the predicted mass flow relies on the {detailed RIAF modelling} and could be reduced, if additional processes, e.g. jet emission, contribute to the continuum emission.
Therefore, this argument should be interpreted carefully.

\subsection{Comparison with X-ray observations}
Hard X-rays are likely emitted in the innermost region around the black hole and allow to investigate the accretion flow.
{Above} $10\,$keV, \citet{2009ApJ...698..528B} found no significant Compton reflection, which is a signature of the inner region of the accretion disc. {However}, their detection of the relativistically broadened fluorescent Fe K$\alpha$ line at $6.4\,$keV, which is usually associated to the inner edge of an accretion disc, indicates an origin of the emission within $\lesssim 45\,r_\text{g}$ of the black hole.
A possible explanation is that Compton reflection might not be the dominant process in creating the X-ray emission in this source.
Alternatively, the broadened iron line component might not be connected to the inner edge of the accretion disc, but produced by optically thick material close to the base of the jets, which would be also the production site of hard X-rays.
In this case the inner edge of the disc could be located even further outside.
If the iron line traces the inner edge of the accretion disc, our evidence for a truncated disc at $r_\text{tr}\gtrsim {26}\, r_\text{g}$ {would also be} consistent with the limits of \citet{2009ApJ...698..528B} and the truncation would help to explain the low Compton reflection observed.

\section{Summary and outlook}
The sub-arcsec resolution SED for the nucleus of NGC 1052, a prototypical LLAGN, does not show the big blue bump associated to the presence of an accretion disc.
In this work, we took advantage of high-angular resolution {observations} to estimate the maximum possible accretion power of a standard accretion disc, corresponding to a mass accretion rate of $\sim 7.1 \times 10^{-6}\,\text{M}_{\sun} \, \text{yr}^{-1}$.
In addition, the sub-arcsec measurements with the VLT reveal a SED dominated by non-thermal emission, of synchrotron origin.
From the sub-arcsec continuum emission we derive a conservative estimate for the rest-frame luminosity.
{This} is higher than the maximum accretion power of a standard accretion disc by one order of magnitude -- a standard accretion disc cannot provide the {required} power {to supply} the continuum emission.

Therefore, we explore the possibility of a truncated accretion disc in order to reconcile the inferred accretion power with the rest-frame luminosity, and find that the maximum mass-light conversion efficiency of the accretion disc scales inversely with the truncation radius.
{For the first time, a truncation of the accretion disc is inferred from high-angular resolution data only.}
In order to provide the necessary accretion power in NGC 1052{,} a thin accretion disc with an accretion rate of $\sim {7.3} \times 10^{-5}\,\text{M}_{\sun} \, \text{yr}^{-1}$ truncated at $r_\text{tr} \gtrsim {26}\, r_\text{g}$ is required.
This estimate is consistent with the results of \citet{2009ApJ...698..528B} obtained from X-ray observations, who find no significant disc reflection in the hard X-ray spectrum and confine the broad Fe K$\alpha$ line emission region to within $\lesssim 45\,r_\text{g}$ of the black hole.

The sub-arcsec resolution sampling of the LLAGN NGC 1052 reveals a continuum flux distribution {that} can be {described} by a broken power law {plus an inverse Compton emission component}.
{With these means}, NGC 1052 does not represent a unique case, instead this seems to be a common {characteristic} within the sub-arcsec resolution LLAGN sample of \citet{2012JPhCS.372a2006F}.
These LLAGNs with presumably jet-dominated spectra \citep{2012AJ....144...11M,2015ApJ...814..139K} will be discussed in a forthcoming paper (Fernández-Ontiveros et al. in prep.).

Up to now the search for a direct link between the Fe K$\alpha$ line broadening and the truncation radius in X-ray binaries has not revealed an unambiguous picture \citep[e.g.][]{2014MNRAS.437..316K,2014arXiv1411.7411P}.
The importance and efficiency {of RIAFs and jets} are subject to an ongoing discussion \citep[e.g.][]{2003MNRAS.343L..99F,2014ARA&A..52..529Y, 2015ApJ...814..139K}.
The influence of additional mechanisms, e.g. {accretion via a corona} or advection into the black hole, is still not well understood.
Therefore, the {new} approach presented in this letter can be applied to other LLAGNs to shed some light on the changes in the internal structure of the accretion disc that are predicted at low accretion rates.

\section*{Acknowledgements}
The authors would like to thank Sera Markoff for providing access to the compact jet--disc model.
{The authors would like to thank the referee for valuable comments which helped to improve the manuscript.}
LR is grateful to Eugene Churazov for useful discussions.
JAFO and AP acknowledge financial support from the Spanish Ministry of Economy and Competitiveness (MINECO) under grant number MEC-AYA2015-53753-P.
KD acknowledges the support of the DFG Cluster of Excellence “Origin and Structure of the Universe” and the Transregio programme TR33 “The Dark Universe”.
This research has made use of NASA's Astrophysics Data System{, the NASA/IPAC Infrared Science Archive,} and of the NASA/IPAC Extragalactic Database (NED).



\bibliographystyle{mnras}



\bsp	
\label{lastpage}
\end{document}